\documentclass[prl,preprint,showpacs,amsmath,amssymb]{revtex4}
\usepackage{enumerate}
\usepackage{color}

\begin{document}
\preprint{YITP-15-32, OU-HET-859}
\title{Geometries from field theories}

\author{Sinya Aoki$^{1,2}$, Kengo Kikuchi$^1$, Tetsuya Onogi$^3$}

\affiliation{
$^1${Yukawa Institute for Theoretical Physics, Kyoto University, Kyoto 606-8502, Japan}\\
$^2${Center for Computational Sciences, University of Tsukuba, Ibaraki 305-8577, Japan}\\
$^3${Department of Physics, Osaka University, Toyonaka, Osaka 560-0043, Japan}
}


\begin{abstract}
We propose a method to define a $d+1$ dimensional  geometry from  a $d$ dimensional quantum  field theory  in  the $1/N$ expansion. We first construct  a $d+1$ dimensional  field theory from the $d$ dimensional one via the gradient flow equation, whose flow time $t$ represents the energy scale of the system such that $t\rightarrow 0$ corresponds to the ultra-violet (UV) while $t\rightarrow\infty$ to the infra-red (IR).  We then define the induced metric from $d+1$ dimensional field operators.    
We show that the metric defined in this way becomes classical in the large $N$ limit, in a sense that quantum fluctuations of the metric are suppressed as $1/N$ due to the large $N$ factorization property.
As a concrete example, we apply our method to the O(N) non-linear $\sigma$ model in two dimensions.
We calculate  the three dimensional induced metric, which is shown to describe  an AdS  space in the massless limit. 
We finally discuss several open issues  in future studies.
\end{abstract}

\maketitle

\noindent
{\em 1. Introduction}\hskip 0.3cm
One of the most surprising and significant findings in field theories and string theories is 
the AdS/CFT (or more generally Gravity/Gauge theory) correspondence\cite{Maldacena:1997re},
which claims a $d$ dimensional conformal field theory is equivalent to some $d+1$ dimensional (super-)gravity theory on the AdS background. After this proposal, there appeared tremendous  number of evidences that this correspondence seems true.
This equivalence is, however, still mysterious and need to be understood, even though the open string/closed string duality may explain it.  

In this paper, we consider such gravity/field theory correspondences from a different point of view, and  propose an alternative method to define a  geometry from a field theory.  
Explicitly, we consider a $d$ dimensional quantum filed theory in the large $N$ expansion, and lift it
to a $d+1$ dimensional one using the gradient flow\cite{Narayanan:2006rf,Luscher:2010iy, Luscher:2009eq,Luscher:2013vga}, where the flow time $t$ becomes an additional coordinate and represents the energy scale of the original $d$ dimensional theory. We then define the induced metric from the field in this $d+1$ dimensional theory. In this way, we define the metric from 
the original $d$ dimensional theory and its scale dependence, and the method is  quite generic and can be applied to all field theories in principle. As shown in the next section, however, the metric defined in this way becomes classical only in the large $N$ limit, thus the use of the large $N$ expansion for field theories may be mandatory.

This paper  is organized as follows. We first propose our method for general field theories in the large $N$ expansion to define the induced metric. We discuss some general properties of this construction.
We next apply the method to the two dimensional O(N) non-linear $\sigma$ model.  We calculate the vacuum expectation value (VEV) of the  three dimensional induced metric in the particular definition, which is shown to 
describe (asymptotically) an AdS space in the massless (UV) limit.   We finally discuss several remaining issues of the method in future studies.  \\

\noindent
{\em 2. Proposal}\hskip 0.3cm
We consider the generic large $N$ field $\varphi^{a,\alpha}(x)$ where $x$ is $d$ dimensional space-time coordinate, $a=1,2,\cdots, $ is the large $N$ index, while $\alpha$ represents other indices such as spinor or vector indices, so that $h_{\alpha\beta}\varphi^{a,\alpha}(x)\varphi^{b,\beta}(x)$ can be made Lorentz invariant by a constant tensor $h_{\alpha\beta}$. We denote the action of this theory $S$.

We first extend the $d$ dimensional field $\varphi^{ a, \alpha}(x)$ to $\phi^{ a, \alpha}(t,x)$ in $d+1$ dimensions, using the gradient flow equation as\cite{Kikuchi:2014rla}
\begin{eqnarray}
\frac{d}{d t} \phi^{a,\alpha}(t,x) &=& - g^{ab}(\phi(t,x))\left. \frac{\delta S}{\delta \varphi^{b,\alpha}(x)}\right\vert_{\varphi \rightarrow\phi} , 
\label{eq:GFE}
\end{eqnarray}
with an initial condition that $\phi^{a,\alpha}(0,x)=\varphi^{a,\alpha}(x)$, 
where $g^{ab}$ is the metric of the space of the large $N$ index.
Since the length dimension of $t$ is 2 and $t\ge 0$, we introduce new variable $\tau =2\sqrt{t}$. (Here a factor 2 makes some latter results simpler. ) Then we denote $d+1$ dimensional coordinate as $z=(\tau,x)\, \in \mathbb {R}^+(=[0,\infty])\times \mathbb{R}^d$ and the field as $\phi^{a,\alpha}(z)$.  

We propose to define a $d+1$ dimensional metric as
\begin{eqnarray}
\hat g_{\mu\nu} (z) &: =& g_{ab}(\phi(z)) h_{\alpha\beta} \partial_\mu \phi^{a,\alpha}(z) \partial_\nu \phi^{b,\beta}(z) ,
\label{eq:metric}
\end{eqnarray}
where a mass dimension of  the constant tensor $h_{\alpha\beta}$\footnote{In general, we may introduce a $z$ dependency tensor $h_{\alpha\beta}(z)$ here, but  we consider the constant case only in this paper.}
must be $-2(1+d_\varphi)$ with $d_\varphi$
being that of $\varphi$, to make the metric dimensionless. This is an induced metric from a $d+1$ dimensional manifold $\mathbb{R}^+\times \mathbb{R}^d$ on a curved space in $\mathbb{R}^N$ with the metric $g_{ab}$.
Using the above definition, we then calculate the expectation values of $g_{\mu\nu}$ and its correlations as
\begin{eqnarray}
\langle \hat g_{\mu\nu}(z) \rangle &:=&  \langle  \hat g_{\mu\nu}(z) \rangle_S , \\
\langle \hat g_{\mu_1\nu_1}(z_1) \hat g_{\mu_2\nu_2}(z_2) \rangle &:=& \langle \hat g_{\mu_1\nu_1}(z_1) \hat g_{\mu_2\nu_2}(z_2) \rangle_S, \\
\langle \hat g_{\mu_1\nu_1}(z_1)\cdots \hat g_{\mu_n\nu_n}(z_n) \rangle &:=& \langle \hat g_{\mu_1\nu_1}(z_1) \cdots \hat g_{\mu_n\nu_n}(z_n) \rangle_S, ~~~~
\end{eqnarray}
where $\langle {\cal O} \rangle_S$ is the expectation values of ${\cal O}(\varphi)$ in $d$ dimensions with the action $S$ as
\begin{eqnarray}
\langle {\cal O} \rangle_S &:=& \frac{1}{Z}\int {\cal D}\varphi \, {\cal O}(\varphi)\, e^{-S}, 
\quad Z:= \int {\cal D}\varphi \,  e^{-S} 
\end{eqnarray}
in the large $N$ expansion. Even though the ``composite" operator $\hat g_{\mu\nu}(z)$ contains a product of two local operators at the same point $z$, $\langle \hat g_{\mu\nu}(z)\rangle $ is finite as long as $\tau\not=0$\cite{Luscher:2011bx}. This is the reason why we define the induced metric in $d+1$ dimensions from $\phi$, not the $d$ dimensional induced metric from $\varphi$, which badly diverges.

Thanks to the large $N$ factorization, 
quantum fluctuations of the metric $\hat g_{\mu\nu}$ are suppressed in the large $N$ limit.
For example, the two point correlation function of $\hat g_{\mu\nu}$ behaves as
\begin{eqnarray}
\langle \hat g_{\mu\nu}(z_1)  \hat g_{\alpha\beta}(z_2) \rangle &=&  
\langle \hat g_{\mu\nu}(z_1)  \rangle \langle\hat g_{\alpha\beta}(z_2) \rangle + O\left(\frac{1}{N}\right),~~
\end{eqnarray}
which shows that the induced metric $\hat g_{\mu\nu}$ is classical in the large $N$ limit, and quantum fluctuations are sub-leading and can be calculated in the $1/N$ expansion.
A use of the $1/N$ expansion here seems important for  the geometrical interpretation of 
the metric $\hat g_{\mu\nu}$.  For  example,  in the large $N$ limit, the VEV of the curvature tensor operator is directly obtained from the VEV of $\hat g_{\mu\nu}$ as in the classical theory.\\

\noindent
{\em 3. An example: O(N) non-linear $\sigma$ model in two dimensions}\hskip 0.3cm
As a concrete example of our proposal, we consider the O(N) non-linear $\sigma$ model in two dimensions, 
whose action is given by
\begin{equation}
S=\frac{1}{2g^2}\int d^2x\, \sum_{a,b=1}^{N-1} g_{ab}(\varphi) \sum_{k=1}^2 \left(\partial_k \varphi^a(x) \partial^k\varphi^b(x)\right),
\end{equation}
where 
\begin{equation}
g_{ab}(\varphi) = \delta_{ab} + \frac{\varphi^a\varphi^b}{1-\varphi\cdot\varphi}, \quad
g^{ab}(\varphi) = \delta_{ab} - \varphi^a\varphi^b 
\end{equation}
with $\varphi\cdot\varphi =\sum_{a=1}^{N-1}\varphi^a\varphi^a$, and the $N$-th component of $\varphi$ is expressed in terms of other fileds as $\varphi^N =\pm \sqrt{1-\varphi\cdot\varphi}$, so that
the metric $g_{ab}$ appears in the action. 
The three dimensional metric $g_{\mu\nu}(z)$ will be extracted from this theory, according to our proposal.\\

\noindent
{\em 3.1 Solution to the gradient flow equation in the large $N$}\hskip 0.3cm
In the previous study\cite{Aoki:2014dxa}, the solution of the gradient flow equation has been obtained in the momentum space as
\begin{eqnarray}
\phi^a(t,p) &=& f(t) e^{-p^2 t} \sum_{n=0}^\infty : X_{2n+1}(\varphi, p,t) :
\end{eqnarray}
 where $X_{2n+1}$ only contains $\varphi^{2n+1}$ terms and is $O(1/N^{2n+1})$.
 The leading order term $X_1$ is given by $X_1^a(\varphi,p,t) = \varphi^a(p)$ with
 \begin{eqnarray}
f(t) &=& \frac{1}{\sqrt{1-2\lambda J(t)}}, \quad
J(t) = \int_0^t ds I(s), 
 \quad I(t)=
 \int \frac{d^2q}{(2\pi)^2} \frac{q^2 e^{-2q^2 t}}{q^2+m^2}, 
\end{eqnarray}
where $\lambda = g^2 N$ is the 't Hooft coupling constant, and $m$ is the dynamically generated mass, which satisfies
\begin{equation}
1= \lambda \int \frac{d^2q}{(2\pi)^2} \frac{1}{q^2+m^2} .
\end{equation}
Introducing the momentum cut-off $\Lambda$, we have
\begin{eqnarray}
f(t) &=&e^{-m^2 t} \sqrt{\frac{\log(1+\Lambda^2/m^2)}{{\rm Ei}\left(-2t(\Lambda^2+m^2)\right)- {\rm Ei}\left(-2t m^2\right)}},~~~
\end{eqnarray}
where  ${\rm Ei}(x)$ is the exponential integral function defined by
${\rm Ei}(-x) = \int d\,x\, e^{-x}/{x}$.
The 2-pt function, which dominate in the large $N$ limit, is calculated as
\begin{eqnarray}
\langle \phi^a(t,x)\phi^b(s,y)\rangle_S &=&  \int  \frac{d^2 q}{(2\pi)^2}  \frac{e^{-q^2 (t+s)} e^{iq(x-y)}}{q^2+m^2} 
\delta_{ab} \frac{\lambda}{N} f(t)f(s)
+O(N^{-2}).
\end{eqnarray}
 
\noindent
{\em 3.2 Induced metric}\hskip 0.3cm 
An  induced metric for this model is given by
\begin{eqnarray}
\hat g_{\mu\nu}(z) := h\, g_{ab}(\phi (z) ) \partial_\mu \phi^a(z) \partial_\nu \phi^b (z) ,
\label{eq:metric_3d}
\end{eqnarray}
where $z = (2\sqrt{t},x)$, and the constant $h$ is introduced so that the mass dimension of the metric operator $\hat g_{\mu\nu}(z)$ is zero.
This is the induced metric from a three dimensional manifold $\mathbb{R}^+\times \mathbb{R}^2$ on the $N-1$ dimensional sphere defined by $\phi^a$. 

The VEV of the metric, $g_{\mu\nu}$,  which does not depend on $x$ due to the translational invariance of the two dimensional O(N) non-linear $\sigma$ model, can easily be calculated in the large $N$ limit as
$g_{i\tau} (\tau) = g_{\tau i} (\tau) = 0$ for $i=1,2$, while
\begin{eqnarray}
g_{ij} (z) &:=& \langle \hat g_{ij} (z ) \rangle = h \langle g_{ab}(\phi) \partial_i\phi^a(t,x) \partial_j\phi^b(t,x) \rangle 
\simeq
\frac{ {h}}{2} \delta_{ij} \lambda f^2(t) I(t ) = \frac{ h}{2} \delta_{ij} \frac{\dot{f}(t)}{f(t)}
\end{eqnarray}
for $i,j=1,2$, 
where  we use an equality that $\dot f(t):=\displaystyle d f(t)/d t = \lambda f(t)^3 I(t)$.  
Furthermore,
\begin{eqnarray}
g_{\tau\tau} (\tau) &=&  \frac{\tau^2 h}{4} \left\langle 
\dot \phi^a(t,x) g_{ab}(\phi(t,x)) \dot \phi^b(t,x) 
\right\rangle ,
\end{eqnarray}
which, after using the gradient flow equation, is  evaluated as
\begin{eqnarray}
g_{\tau\tau} (\tau) 
 &\simeq& \frac{\tau^2  h}{4} \left[ \langle \nabla^2\phi\cdot  \nabla^2\phi\rangle - \langle \phi\cdot  \nabla^2\phi\rangle^2\right]  
 =-\frac{\tau  h}{4}\frac{d}{d \tau} \left(\frac{\dot f}{f}\right) .
\end{eqnarray}
Thus the expectation values of the induced metric turns out to be diagonal as
\begin{equation}
g_{\mu\nu} = \left(\begin{array}{ccc}
B(\tau) & 0 & 0 \\
0 & A(\tau) & 0 \\
0 & 0 & A(\tau) \\
\end{array}
\right)
\end{equation}
where we define
\begin{equation*}
A(\tau) =  \displaystyle \frac{ h}{2}\frac{\dot f(t)}{f(t)}\Bigr\vert_{t=\tau^2/4},\
B(\tau) =   -\frac{\tau}{2} A_{,\tau},
\end{equation*}
and $f_{,\tau}$ means the derivative of $f$ with respect to $\tau$.
This $A(\tau)$, hence also $ B(\tau)$, is finite in the $\Lambda\rightarrow\infty$ limit as
\begin{eqnarray}
A(\tau) &=&- \frac{m^2  {h}}{2}\left[1 +\frac{e^{-\tau^2 m^2/2}}{{\rm E_i}(-\tau^2m^2/2) m^2\tau^2/2 }\right].
~~
\label{eq:A}
\end{eqnarray}

From the metric, we can calculate the VEV of composite operators such as the Einstein tensor $G_{\mu\nu}(\hat g)$ as $\langle G_{\mu\nu}(\hat g)\rangle = G_{\mu\nu}(\langle\hat g\rangle)$,  thanks to the factorization in the large $N$ limit.

After the little algebra, we obtain
\begin{equation}
G_{\tau\tau} =\displaystyle \frac{A_{,\tau}^2}{4 A^2},\quad 
G_{ij} =\displaystyle \delta_{ij} \left[ \frac{A_{,\tau\tau}}{2B} - \frac{A_{,\tau} B_{,\tau}}{4B^2} - \frac{A_{,\tau}^2}{4AB}\right],
\end{equation}
and $G_{i\tau} = G_{\tau i} = 0$. \\

\noindent
{\em 3.3 Massless limit and AdS space} \hskip 0.3cm

We consider the massless limit ($m\rightarrow 0$),
where $A$ and its derivatives  are given by
\begin{eqnarray}
A &\simeq & -\frac{1}{\tau^2} \frac{h}{\log (m^2)} \left[1+O\left( \frac{1}{\log(m^2)}\right)\right], 
\end{eqnarray}
$A_{,\tau} \simeq -2A/\tau$ and
$A_{,\tau\tau} \simeq 6 A/\tau^2$. 
We here use the expansion ${\rm E_i}(-x) =\log x +\gamma+ \sum_{n=1}^\infty (-x)^n/(n\cdot n!)$.
In order to have positive and finite $g_{\mu\nu}$ in the massless limit, we take $h=-R_0^2 \log (m^2 R_0^2)$, where  a 
mass dimension of the constant $R_0$ is $-1$.
We thus obtain
$g_{\tau\tau}  = \displaystyle\frac{R_0^2}{\tau^2}$,  $g_{ij}  = \displaystyle\delta_{ij} \frac{R_0^2}{\tau^2}$,
so that 
\begin{equation}
ds^2 = \frac{R_0^2}{\tau^2}\left[  d\tau^2 + (d\vec{x})^2 \right],
\end{equation}
which describes the Euclidean AdS space. Indeed, the Einstein tensor reads
\begin{equation}
G_{\mu\nu} = -\Lambda_0 g_{\mu\nu}, \qquad
\Lambda_0 = - \frac{1}{R_0^2} , 
\end{equation} 
which give the negative cosmological constant $\Lambda_0$.
It is interesting to see that the AdS geometry is realized for the conformal field theory defined in the massless limit, which corresponds to the UV fixed point of the theory.
\\

\noindent
 {\em 3.4 Metric in UV and IR limits} \hskip 0.3cm
 In the short distance (UV) limit  ($m\tau \rightarrow 0$), we have
$A \simeq  -h/[\tau^2 \log (m^2\tau^2)]$, 
so that
$\displaystyle g_{\tau\tau} \simeq - h/[\tau^2 \log (m^2\tau^2)]$ and 
$\displaystyle g_{ij} \simeq - \delta_{ij}h/[\tau^2  \log (m^2\tau^2)]$.

As briefly discussed for generic cases,
had  we defined the $d$ dimensional metric directly from the $d$ dimensional field theory as
$\hat g_{ij}^d(x) := g_{ab}( \varphi(x)) \partial_i \varphi^a(x)  \partial_j \varphi^b(x)$,
the VEV of $\hat g_{ij}^d(x)$ would become UV divergent due to the short distance singularity of $\varphi^a (x) \varphi^b (y)$ at $x\rightarrow y$.
In contrast,  the $d+1$ dimensional metric $\hat g_{\mu\nu}$ defined from the $d$ dimensional field theory via eq.~(\ref{eq:metric}) is free from UV divergence, since the flowed field $\phi (t,x)$ and any local composite operators constructed from it 
are expected to be finite as long as $t\sim\tau^2$ is non-zero\cite{Luscher:2011bx,Aoki:2014dxa,Makino:2014sta,Makino:2014cxa}.
This is the reason why we employ flowed field and the induced metric is defined on $d+1$ dimensions, not on $d$ dimensions.  Consequently, the classical metric, $g_{\mu\nu}$, is UV finite in our proposal, as it should be. 
Note that the UV divergence presented in the two dimensional O(N) non-linear $\sigma$ model, for example,  appears in the $m\tau \rightarrow 0$ limit as $g_{\mu\nu} \sim 1/(\tau^2\log m^2\tau^2)$.

In this limit, the ``effective" cosmological constant is given as 
\begin{equation}
\Lambda_0^{\rm eff} = -\frac{\log (m^2\tau^2)}{R_0^2\log(m^2R_0^2)},
\end{equation} 
where  non-conformal natures of the original two dimensional asymptotic-free field theory appear in its $\log \tau^2$ dependence.

In the $m\tau\rightarrow\infty$ (IR) limit, on the other hand,
$A \simeq  h/\tau^2 $, 
which give
$\displaystyle g_{\tau\tau} \simeq h/\tau^2$ and
$\displaystyle g_{ij} \simeq  \delta_{ij}h/\tau^2 $.
Since $\Lambda_{0}^{\rm eff} = [R_0^2\log(m^2R_0^2)]^{-1}$ in this limit,  
the theory becomes asymptotically  AdS ($\Lambda_0^{\rm eff} < 0$) if  $\log(m^2 R_0^2) < 0$.
This result  looks rather non-trivial,
since  the massive theory
is expected naively to become trivially conformal due to decouplings of all massive modes in the IR limit. 

Assuming the Einstein equation,  $G_{\mu\nu} =8\pi G T_{\mu\nu}$, we can define the energy momentum tensor $T_{\mu\nu}$.  In both UV and IR limits, we have
$T_{\mu\nu}=\delta_{\mu\nu}/(8\pi G \tau^2)$, which does not depend on $m$ and therefore holds even at $m=0$.
As a consequence,  $T_{\mu\nu}^{\rm matter}$ vanishes in both UV and IR limits, where we define $T_{\mu\nu}^{\rm matter}:=T_{\mu\nu} + g_{\mu\nu}\Lambda_0/(8\pi G)$. \\

\noindent
{\em 4. Discussions}\hskip 0.3cm 
In our proposal, a correspondence between a  geometry and a field theory is explicit by construction,
and 
the method proposed here can be applied to an arbitrary quantum field theory,
as long as  the large $N$ expansion is employed.
If the theory is solvable in the large $N$ limit, the VEV of $\hat g_{\mu\nu}$ can be calculated exactly.
Let us discuss  the remaining issues  which should be investigated in future studies. 

First, we do not  have so far a full dictionary  to interpret a quantum field theory in terms of the corresponding  metric operator and vice versa.  For the two dimensional O(N) non-linear $\sigma$ model,   the dynamically generated mass  $m$ can be extracted 
from the asymptotic behavior of the metric that $g_{\mu\nu}\sim 1/[\tau^2\log(m^2\tau^2)]$ as $m\tau\rightarrow 0$. However, this gives only a partial knowledge of the field theory. It would be nice if we could determine a whole structure of  the matter content based on 
a gravity action which gives the same Einstein equation we assumed. 

Secondly,  since our method can be applied to a large class of quantum field theories,
one should investigate what kind of geometry emerges from
various large $N$ models other than the two dimensional O(N) non-linear $\sigma$ model, including conformal theories.
One possible direction is to introduce a source field in the original theory to break the translational invariance 
to generate geometries with nontrivial $z$ dependence. 
In particular, it  would be a challenge to find the field theory set-up which induces the black hole geometry. 

Thirdly,  the fluctuations around the background geometry should be studied.
In principle, we can calculate an arbitrary correlation function for the metric $\hat g_{\mu\nu}$ including the 
quantum fluctuation of the metric in the $1/N$ expansion.
In practice, however, calculations in the next leading order  become much more complicated than those in the large $N$ limit\cite{Aoki:2014dxa}.
Although no action is explicitly given for $\hat g_{\mu\nu}$ in our approach, one may effectively define the quantum theory of the metric in this way. 
It would be interesting to investigate whether this quantum theory  is renormalizable (or even UV finite) or not, in contrast to the unrenormalizable quantum theory of the Einstein gravity.
Furthermore, it is also interesting to calculate an effective action for the composite operator of
$\hat g_{\mu\nu}$.

Since the equation (\ref{eq:GFE}) is not a unique way to define the flow equation, and thus $d+1$ dimensional field $\phi$ from $d$ dimensional $\varphi$, a dependence  of the induced  metric on the flow equation should be investigated. 

Finally, in the case of gauge theories, simple choices for 
the induced metric may be
\begin{eqnarray}
\hat g_{\mu\nu}(z) &:=& h \sum_{i,j=1}^d{\rm Tr}\ D_\mu F_{ij}(z) D_\nu F^{ij}(z), \\
\hat g_{\mu\nu}(z) &:=& h \sum_{\alpha=0}^d {\rm Tr}\ F_{\mu\alpha}(z)F_{\nu}{}^{\alpha}(z),
\end{eqnarray}
both of which are invariant under the $\tau$-independent gauge transformation\cite{Luscher:2010iy,Luscher:2009eq,Luscher:2013vga}. Here
 $D_i$ ($i=1,\cdots,d$) is the covariant derivative in $d$ dimensions while $D_\tau :=\partial_\tau$,  and then the field strength is given as $F_{\mu\nu}:=[D_\mu,D_\nu]$ .  
It will be interesting to calculate the induced metric form 
the large $N$ gauge theory in two dimensions ('t Hooft model)\cite{'tHooft:1974hx} in our method.

\bigskip

S.A. would like to thank Drs.  N. Ishibashi, S. Sugimoto, T. Takayanagi and Y. Yokokura for their useful comments. 
 We also thank Dr. S. Yamaguchi for his comment on the AdS metric.
This work was supported by Grant-in-Aid for JSPS Fellows Grant Number 25$\cdot$1336 and by the Grant-in-Aid of the Japanese Ministry of Education (Nos. 25287046,26400248), by MEXT SPIRE and JICFuS.

\end{document}